\documentclass[twocolumn,showpacs,amsmath,prb,final]{revtex4}
\usepackage{graphicx}% Include figure files
\usepackage{dcolumn}% Align table columns on decimal point
\usepackage[sort&compress]{natbib}
\usepackage{subfigure}
\usepackage{ifpdf}
\usepackage{bm}
\usepackage{amsmath}

\begin{document}
\title{Nanoscale superconductors: quantum confinement and spatially
dependent Hartree-Fock potential}

\author{Yajiang Chen}
\affiliation{Departement Fysica, Universiteit Antwerpen,
Groenenborgerlaan 171, B-2020 Antwerpen, Belgium}
\author{M. D. Croitoru}
\affiliation{Departement Fysica, Universiteit Antwerpen,
Groenenborgerlaan 171, B-2020 Antwerpen, Belgium}
\author{A. A. Shanenko}
\email{arkady.shanenko@ua.ac.be}\affiliation{Departement
Fysica, Universiteit Antwerpen, Groenenborgerlaan 171, B-2020
Antwerpen, Belgium}
\author{F. M. Peeters}
\email{francois.peeters@ua.ac.be}
\affiliation{Departement Fysica, Universiteit Antwerpen,
Groenenborgerlaan 171, B-2020 Antwerpen, Belgium}

\date{\today}
\begin{abstract}
It is well-known that in bulk, the solution of the Bogoliubov-de Gennes
equations is the same whether or not the Hartree-Fock term is included.
In this case the Hartree-Fock potential is position independent and, so,
gives the same contribution to both the single-electron energies and
the Fermi level (the chemical potential). Thus, the single-electron
energy measured from the Fermi level (it controls the solution) stays
the same. It is not the case for nanostructured superconductors, where
quantum confinement breaks the translational symmetry and results in
a position dependent Hartree-Fock potential. Now the contribution of
the Hartree-Fock mean field to the single-electron energies depends
on the relevant quantum numbers. Moreover, the single-electron wave
functions can be influenced by the presence of this additional spatially
dependent field. We numerically solved the Bogoliubov-de Gennes equations
with the Hartree-Fock term for a clean metallic nanocylinder and found
a shift of the curve representing the thickness-dependent oscillations
of the critical temperature (the energy gap, the order parameter etc.)
to larger diameters. Though the difference between the superconducting
solutions with and without the Hartree-Fock interaction can, for some
diameters, be very significant, the above mentioned shift is less than
typical metallic unit-cell dimensions and, so, has no practical worth.
This allows one to significantly simplify the problem and, similar to
bulk, ignore the Hartree-Fock potential when solving the Bogoliubov-de
Gennes equations in the nano-regime.
\end{abstract}
\maketitle

\section {Introduction}
\label{int}

Advances in nanofabrication technology resulted recently in high-quality
metallic superconducting ultrathin nanofilms~\cite{film1,film2,film3}
and nanowires.~\cite{wire1,wire2,wire3,wire4} In most samples the
electron mean free path was estimated to be about or larger than the
nanofilm/nanowire thickness.~\cite{film2,wire1,wire4} In this case the
effects of the transverse quantization are not shadowed by impurity
scattering and, hence, the conduction band splits up into a series of
single-electron subbands resulting from the quantized transverse modes.
This will have a pronounced effect on the superconducting properties
(see, for instance, Refs.~\onlinecite{sh1} and \onlinecite{sh2} and
references therein). Notice that high-quality nanofilms do not exhibit
significant indications of defect- or phase-driven suppression of
superconductivity (see discussion in Ref.~\onlinecite{film2}). For high%
-quality nanowires the phase-fluctuation effects were shown to seriously
influence the superconducting state only in narrowest aluminum specimens
with width $\approx 5-8\,{\rm nm}$.~\cite{wire1,wire4,arut} Thus, the
transverse quantum confinement is the major mechanism governing the
superconducting properties in this case. Therefore, it is timely to
study in a more detail a clean nanoscale superconductor in the presence
of quantum confinement.

Quantum confinement breaks the translational symmetry and, so, the
superconducting order parameter becomes position dependent. The well-%
known BCS ansatz for the ground state wave function is not applicable in
this case, and the Bogoliubov-de Gennes (BdG) equations are a relevant tool
to investigate equilibrium superconducting properties. Recent numerical
studies of the BdG equations for nanofilms~\cite{sh1} and nanowires~%
\cite{sh2,han,grig} show that the transverse quantum confinement has a
substantial impact on the superconducting solution. However, the BdG
equations investigated in Refs.~\onlinecite{sh1}, \onlinecite{sh2},
\onlinecite{han} and \onlinecite{grig}, were solved without the Hartree-%
Fock (HF) potential. The reason is that in bulk, the superconducting
solution is not sensitive to the HF term in the BdG equations~\cite{fett},
and one can assume that a similar conclusion holds for the broken
translational symmetry. However, at present there is no detailed
investigations on this subject and, so, such a study is needed.

In the bulk BdG equations, the HF potential is not spatially dependent and,
so, it produces the same contribution to all single-electron energies, with
no dependence on the relevant quantum numbers. Hence, the Fermi level (the
chemical potential) acquires the same contribution, as well, and the single-%
electron energies measured from the Fermi level are not changed. It is well-%
known that the BdG equations are derived within the grand canonical formalism
and, so, the electron energies appearing in the basic expressions absorb the
chemical potential. As a result, the superconducting solution is insensitive
to the HF potential. The situation is different in the presence of quantum
confinement. The translational symmetry is now broken, the HF mean field is
position dependent, and, so, its contribution to the single-electron energies
is a function of the relevant quantum numbers. Furthermore, the single-%
electron wave functions themselves are influenced by the presence of the HF
field, i.e., an additional spatially-dependent potential. Therefore, one can
expect that the HF term in the BdG equations can change the superconducting
solution in the presence of quantum confinement. It is of importance to
clarify to what extent this will be through. In particular, this concerns
the thickness-dependent oscillations (i.e., quantum-size oscillations) of
the superconducting properties typical of high-quality nanofilms and
nanowires.~\cite{film1,film2,sh1,sh2}.

In the present work, based on the particular case of a superconducting clean
metallic nanowire of the cylindrical form, we compare the superconducting
numerical solutions of the BdG equations with and without the HF potential.
We find that these solutions are indeed different and, for some nanocylinder
diameters, this can be very significant. However, a more close look reveals
that this difference is actually expressed in a small shift of the curve
representing thickness-dependent oscillations of the critical temperature
(or other important superconducting quantities) up to larger diameters.
The shift is less than typical metallic unit-cell dimensions and, so, can
be ignored.

The paper is organized as follows. In Sec.~\ref{sec2} the formalism of the
BdG equations is outlined, with the focuser on the features related to the
cylindrical confining geometry. In addition, the Anderson approximate
solution to the BdG equations is discussed in this section. To check the
effect of the HF term on the single-electron wave functions, the Anderson
solution is constructed by assuming that these wave functions do not change
in the presence of the HF interaction. In Sec.~\ref{sec3} numerical results
of the BdG equations with and without the HF term are discussed. Based on
the Anderson solution, here we also investigate the effect of the HF term
on the single-electron wave functions.

\section{Bogoliubov-de Gennes equations and Anderson's recipe}
\label{sec2}

We focus on the basic superconducting properties of a metallic clean
cylindrical nanowire (with diameter $D=2R$ and length $L$) in the
quantum-size regime when the transverse quantization of the single-%
electron spectrum is of importance. In the presence of quantum
confinement the translational invariance is broken, and the order
parameter appears to be position-dependent, i.e., $\Delta({\bf r})$.
It is well-known that the BdG equations are a common and useful
approach to investigate such a situation. Generally, these equations
can be represented as follows:
\begin{subequations}
\begin{align}
&E_{\nu}|u_{\nu}\rangle=\hat{H}_e|u_{\nu}\rangle + \hat{\Delta}
|v_{\nu}\rangle,\label{BdG1}\\
&E_{\nu}|v_{\nu}\rangle=\hat{\Delta}^{\ast}|u_{\nu}\rangle
-\hat{H}^{\ast}_e|v_{\nu}\rangle,\label{BdG2}
\end{align}
\end{subequations}
where $E_{\nu}$ stands for the quasiparticle energy, $|u_{\nu}\rangle$
and $|v_{\nu}\rangle$ are the particle-like and hole-like ket vectors.
In the clean limit the single-electron Hamiltonian in Eqs.~(\ref{BdG1})
and (\ref{BdG2}) is of the form [for zero magnetic field, ${\bf A}=0$]
\begin{equation}
\hat{H}_e=\hat{H}^{\ast}_e=\frac{{\hat{\bf p}}^2}{2m_e} +
\Phi_{HF}({\hat{\bf r}}) + V_{\rm conf}({\hat{\bf r}})- E_F,
\label{He}
\end{equation}
with ${\hat {\bf r}}$ and ${\hat {\bf p}}$ the position and momentum
operators, $E_F$ the Fermi level, $m_e$ the electron band mass~(set to the
free electron mass), $V_{\rm conf}({\bf r})$ the confining interaction,
and $\Phi_{HF}({\bf r})$ the HF potential. In bulk the confining interaction
can be neglected and we arrive at the usual BCS picture based on plane waves.
Below we adopt the simplest choice of the confining interaction potential:
zero inside and infinite outside the wire. The gap-operator $\hat{\Delta}$
in Eqs.~(\ref{BdG1}) and (\ref{BdG2}) is related to the order parameter by
$\hat{\Delta} =\Delta({\hat {\bf r}})$.

As a mean-field theory, the BdG equations are solved in a self-consistent
manner with the self-consistency relations given by
\begin{subequations}
\begin{align}
&\Delta({\bf r}) = g\sum_{\nu\in\mathcal{C}}\langle {\bf r}|u_{\nu}
\rangle\langle v_{\nu}|{\bf r}\rangle \bigl[1-2f_{\nu}\bigr],
\label{self1} \\
&\Phi_{HF}({\bf r}) = -g\sum_{\nu}\Bigl[|\langle {\bf r}|u_{\nu}
\rangle|^2 f_{\nu}+|\langle{\bf r}|v_{\nu}\rangle|^2(1-f_{\nu})
\Bigr],
\label{self2}
\end{align}
\end{subequations}
where $g>0$ is the coupling constant, $f_{\nu}=1/(e^{\beta E_{\nu}}+1)$ is
the Fermi function [$\beta=1/(k_BT)$ with $T$ the temperature and $k_B$ the
Boltzmann constant]. In Eq.~(\ref{self1}) $\mathcal{C}$ indicates the set
of quantum numbers corresponding to the single-electron energy $\xi_{\nu}$
(measured from the Fermi level) located in the Debye window $\xi_{\nu
\in\mathcal{C}} \in [-\hbar \omega_D,\hbar \omega_D]$~($\omega_D$ is the
Debye frequency), where $\xi_{\nu}$ absorbs the HF potential, i.e.,
\begin{equation}
\xi_{\nu}=\langle u_{\nu}|\hat{H}_e|u_{\nu}\rangle + \langle v_{\nu}|
\hat{H}^{\ast}_e|v_{\nu}\rangle.
\label{xi}
\end{equation}
The cut-off in Eq.~(\ref{self1}) is known~\cite{degen} to be a payment for
using a simplified delta-function approximation for the electron-electron
interaction. Such a regularization is not needed in Eq.~(\ref{self2}). For
our confining interaction (i.e., zero inside and infinite outside) we have
\begin{equation}
\langle {\bf r}|u_{\nu}\rangle\Bigl|_{{\bf r}\in S}=\langle {\bf r}|
v_{\nu}\rangle\Bigl|_{{\bf r}\in S}=0
\label{bound}
\end{equation}
at the sample surface, i.e., ${\bf r}\in S$. Periodic boundary conditions
with unit cell $L$ can be applied in the direction parallel to the nanowire.

The Fermi level (i.e., the chemical potential) is determined from
\begin{equation}
n_e =\frac{2}{\pi R^2 L}\sum_{\nu}\Bigl[\langle
u_{\nu}|u_{\nu}\rangle f_{\nu} + \langle v_{\nu}|v_{\nu}\rangle
(1-f_{\nu})\Bigr],
\label{ne}
\end{equation}
where $n_e$ is the mean electron density. We use the BdG equations in
the parabolic band approximation and, so, as discussed in Ref.~%
\onlinecite{sh1}, an effective Fermi level should be introduced, to
recover the correct period of the quantum-size oscillations. For
aluminum (the aluminum parameters are used below) $E_F=0.9\,{\rm eV}$
for $D\approx 10\,{\rm nm}$~(see Ref.~\onlinecite{sh2}). For $D \sim
1-2\,{\rm nm}$, $E_F$ shifts systematically from this value up, due
to Eq.~(\ref{ne}).

Due to the chosen confining geometry, it is convenient to use
cylindrical coordinates $\rho, \varphi$ and $z$.  In this case the
order parameter (the anomalous pairing potential) and HF mean field
(the normal potential) depend only on the transverse coordinate, i.e.,
$\Delta(\rho)$ and $\Phi_{HF}(\rho)$, and $\langle {\bf r}|u_{\nu}
\rangle$ and $\langle{\bf r}|v_{\nu}\rangle$ are represented in the
form $(\nu =\{j,m,k\})$
\begin{equation}
\left(
\begin{array}{c}
\langle {\bf r}|u_{jmk}\rangle\\
\langle {\bf r}|v_{jmk}\rangle
\end{array}
\right)=\frac{e^{\imath m\varphi}}{\sqrt{2\pi}}
\frac{e^{\imath k z}}{\sqrt{L}}
\left(\begin{array}{c}
u_{jmk}(\rho)\\
v_{jmk}(\rho)
\end{array}
\right),
\label{factor}
\end{equation}
with $j$ controlling the number of nodes in the transverse direction,
$m$ the azimuthal quantum number, and $k$ the wave vector of the
quasi-free electron motion along the nanocylinder. Inserting Eq.~%
(\ref{factor}) into Eqs.~(\ref{BdG1}) and (\ref{BdG2}), we recast
the BdG equations as
\begin{subequations}
\begin{align}
\bigl[E_{jmk}- {\cal L}_{\rho} - \Phi_{HF}(\rho)\bigr]\,u_{jmk}(\rho)=
\Delta(\rho)v_{jmk}(\rho),\label{BdGA}\\
\bigl[E_{jmk} + {\cal L}_{\rho} + \Phi_{HF}(\rho)\bigr]\,v_{jmk}(\rho)=
\Delta(\rho)u_{jmk}(\rho),\label{BdGB}
\end{align}
\end{subequations}
where $\Delta(\rho)$ is real, and
\begin{equation}
{\cal L}_{\rho}=-\frac{\hbar^2}{2m_e}\Bigr(\frac{\partial^2}{\partial
\rho^2}+ \frac{1}{\rho}\frac{\partial}{\partial\rho} -
\frac{m^2}{\rho^2} - k^2\Bigl)-E_F.
\end{equation}
The self-consistency relations can be rewritten as,
\begin{subequations}
\begin{align}
&\Delta(\rho)=\frac{g}{2\pi L}\!\sum_{jmk \in C} u_{jmk}(\rho) v_{jmk}(\rho)
\bigl[1-2f_{jmk}\bigr],
\label{selfA}\\
&\Phi_{HF}(\rho)\!=\nonumber\\
&-\frac{g}{2\pi L}\!\sum_{jmk}\Bigl[u^2_{jmk}(\rho) f_{jmk}
+v^2_{jmk}(\rho)\bigl(1-f_{jmk}\bigr)\Bigr],\label{selfB}
\end{align}
\end{subequations}
with $u_{jmk}(\rho)$ and $v_{jmk}(\rho)$ real. To numerically solve Eqs.~%
(\ref{BdGA}) and (\ref{BdGB}), we expand the transverse particle-like and
hole-like wave functions as
\begin{equation}
\left(
\begin{array}{c}
u_{jmk}(\rho)  \\
v_{jmk}(\rho)
\end{array}
\right) =\sum\limits_J
\left(
\begin{array}{c}
u_{jmk,J}\\[1mm]
v_{jmk,J}
\end{array}
\right)\,\vartheta_{\!Jm}(\rho),
\label{series}
\end{equation}
with
\begin{equation}
\vartheta_{\!Jm}(\rho)=\frac{\sqrt{2}}{R{\cal J}_{m+1}(\alpha_{Jm})}{\cal
J}_m(\alpha_{Jm}\frac{\rho}{R}),
\label{singlewave}
\end{equation}
where ${\cal J}_m(x)$ is the Bessel function of the first kind of the $m$-%
order, and $\alpha_{Jm}$ is the $J$th zero of this function. This allows one
to convert Eqs.~(\ref{BdGA}) and (\ref{BdGB}) into a matrix form. Then,
a numerical solution can be obtained by diagonalizing the corresponding
matrix, and self-consistency is reached by iterating Eqs.~(\ref{selfA})
and (\ref{selfB}). One should keep in mind that~\cite{degen} $\langle
u_{\nu}|u_{\nu}\rangle + \langle v_{\nu}| v_{\nu}\rangle = 1$ and, so,
\begin{equation}
\int\limits_{0}^{R}{\rm d}\rho \,\rho [u^2_{jmk}(\rho) + v^2_{jmk}
(\rho)] = 1.
\label{norm}
\end{equation}
As seen, the vector $(u_{jmk,J};v_{jmk,J})^{\rm T}$~($J = 0,1, \ldots$) is
normalized.

%%%%%%%%%%%%%%%%%%%%%%%%%%%%%%%%%%%%%%%%%%%%%%%%%%%%%%%%%%%%%%%%%%%%%%%
\begin{figure*}
\resizebox{1.5\columnwidth}{!}
{\rotatebox{0}{\includegraphics{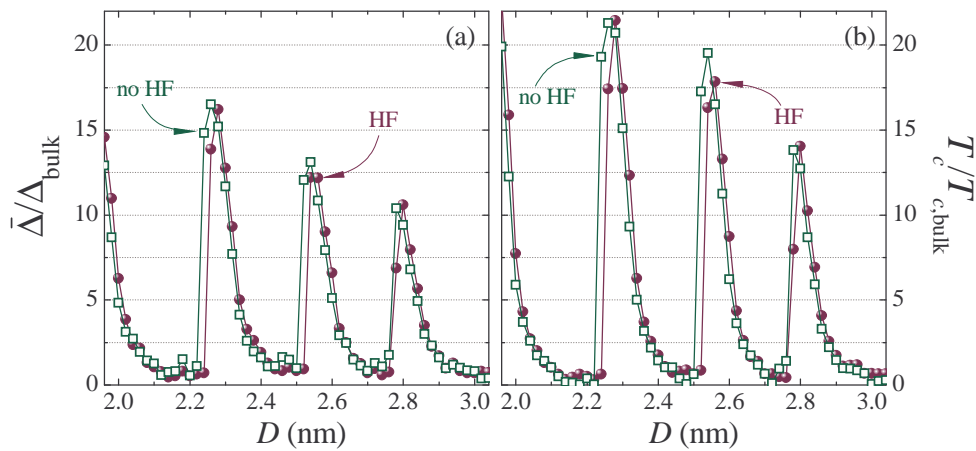}}}
\caption{(a) The spatially averaged superconducting order parameter
$\bar{\Delta}/\Delta_{\rm bulk}$ and (b) critical temperature $T_c/
T_{c,{\rm bulk}}$ versus the nanowires diameter $D$ as calculated
from the BdG equations (\ref{BdGA}) and (\ref{BdGB}) at zero
temperature.}
\label{fig1}
\end{figure*}
%%%%%%%%%%%%%%%%%%%%%%%%%%%%%%%%%%%%%%%%%%%%%%%%%%%%%%%%%%%%%%%%%%%%%%%

In addition to the above procedure, below we use the Anderson approximate
solution, as well~\cite{ander}. Within this approximation, instead of the
expansion given by Eq.~(\ref{series}), it is assumed that
\begin{equation}
u_{jmk}(\rho) = {\cal U}_{jmk}\,\vartheta_{jm}(\rho),\; v_{jmk}(\rho) =
{\cal V}_{jmk}\,\vartheta_{jm}(\rho).
\label{and}
\end{equation}
Equation~(\ref{and}) means that we seek a minimum of the BdG thermodynamic
functional in the subspace of $u_{jmk}(\rho)$ and $v_{jmk}(\rho)$
proportional to the eigenfunctions of ${\cal L}_{\rho}$. Notice that it is
possible to deal with Anderson's recipe, invoking the eigenfunctions of
${\cal L}_{\rho}+\Phi_{HF}(\rho)$. However, below we are interested in Eq.~%
(\ref{and}) because it helps to clarify how a change in the single-electron
wave functions due to the HF potential, can contribute to the problem of
interest. To be accurate, the Anderson approximation should be based on
the true single-electron wave functions. We recently found that in this
case the error in Anderson's solution for $D \lesssim 2-3\,{\rm nm}$ is
less than one-two percents.~\cite{sh3} Hence, comparing the results of
numerically solving Eqs.~(\ref{BdGA}) and (\ref{BdGB}) with the data based
on Eq.~(\ref{and}), we can reach unambiguous conclusions about the role of
the changes in the single-electron wave functions due to the HF interaction.
As follows from Eq.~(\ref{and})~[see, for instance, Ref.~\onlinecite{sh3}],
the Anderson-approximation results in the BCS-like self-consistent equation
\begin{equation}
\Delta_{j'm'}= -\frac{1}{2}\!\sum\limits_{jmk \in C}
\frac{g_{j'm',jm}\;\;\Delta_{jm}}{\sqrt{\xi^2_{jmk}+\Delta^2_{jm}}}
\bigl[1-2f_{jmk}\bigr],
\label{gapAnder}
\end{equation}
with
\begin{equation}
\Delta_{jm}=\int\limits_0^R\!\!{\rm d}\rho\,\rho\,\vartheta^2_{jm}(\rho)
\;\Delta(\rho)
\label{djm}
\end{equation}
and the interaction-matrix element given by
\begin{equation}
g_{j'm',jm}=-\frac{g}{2\pi L}\int\limits_0^R\!\!{\rm d}\rho
\,\rho ~\vartheta_{j'm'}^2(\rho)~\vartheta_{jm}^2(\rho).
\label{int_matrix}
\end{equation}
For the single-electron energy appearing in Eq.~(\ref{gapAnder}) we have
\begin{equation}
\xi_{jmk}=\frac{\hbar^2}{2m_e}\Bigl[\frac{\alpha_{jm}^2}{R^2}+
k^2\Bigr]+\Phi_{jm} - E_F,
\label{xiAnder}
\end{equation}
where
\begin{equation}
\Phi_{jm} = \int\limits_0^R\!\!{\rm
d}\rho\,\rho\,\vartheta^2_{jm}(\rho)\;\Phi_{HF}(\rho).
\label{HFAnder}
\end{equation}
Inserting Eq.~(\ref{self2}) into Eq.~(\ref{HFAnder}), one obtains
\begin{equation}
\Phi_{j'm'} = \frac{1}{2}\sum\limits_{jmk}
g_{j'm',jm}\;\Bigl[1 - \frac{\xi_{jmk}(1-2f_{jmk})}{\sqrt{
\xi^2_{jmk}+\Delta^2_{jm}}}\Bigr].
\label{HFAnder1}
\end{equation}
We should not forget about $E_F$ appearing in the single-electron energy
given by Eq.~(\ref{xiAnder}). It is fixed through Eq.~(\ref{ne}) that is
now of the form
\begin{equation}
n_e =\frac{1}{\pi R^2 L} \sum\limits_{jmk}\Bigl[1 -
\frac{\xi_{jmk}(1-2f_{jmk})}{\sqrt{\xi^2_{jmk}+\Delta^2_{jm}}}
\Bigr].
\label{neAnder}
\end{equation}
Thus, in the Anderson approximation introduced by Eq.~(\ref{and}), one
needs to solve Eqs.~(\ref{gapAnder}) and (\ref{HFAnder1}), keeping
Eq.~(\ref{neAnder}). As already mentioned above, comparing a numerical
solution of Eqs.~(\ref{BdGA}) and (\ref{BdGB}) with the solution based
on Anderson's recipe, we can check the effect of the HF interaction on
the single-electron wave functions.

\section{Numerical results}
\label{sec3}

%%%%%%%%%%%%%%%%%%%%%%%%%%%%%%%%%%%%%%%%%%%%%%%%%%%%%%%%%%%%%%%%%%%%%%%
\begin{figure*}
\resizebox{2.0\columnwidth}{!}
{\rotatebox{0}{\includegraphics{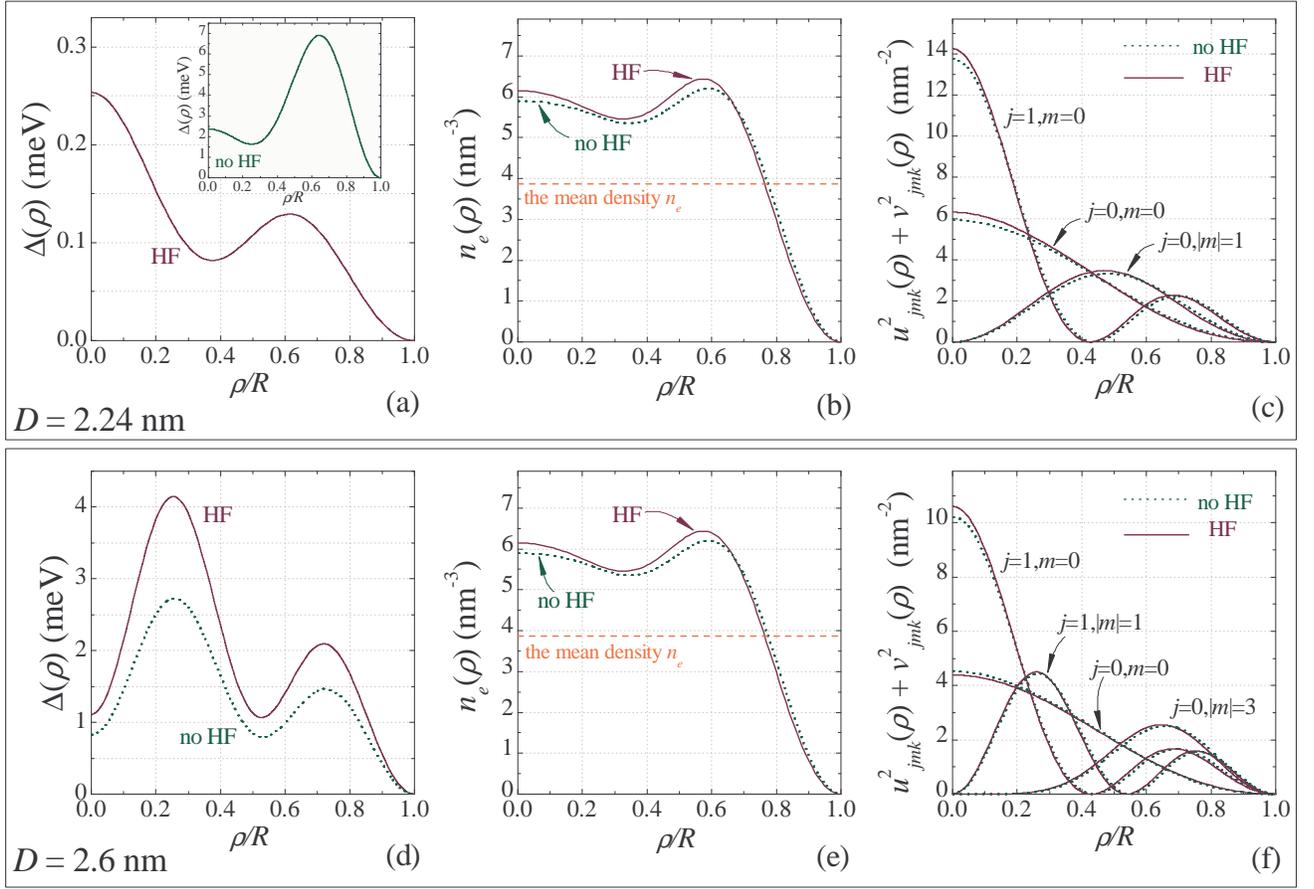}}}
\caption{Upper panel: (a) the superconducting order parameter $\Delta
(\rho)$, (b) the local density $n_e(\rho)$  and (c) the distribution
$u^2_{jmk}(\rho) + v^2_{jmk}(\rho)$ versus $\rho/R$ for $D=2.24\,{\rm
nm}$. The lower panel: the same but for $D=2.6\,{\rm nm}$. The results
for the truncated (no HF) and full (HF) BdG equations are plotted.}
\label{fig2}
\end{figure*}
%%%%%%%%%%%%%%%%%%%%%%%%%%%%%%%%%%%%%%%%%%%%%%%%%%%%%%%%%%%%%%%%%%%%%%%

In this section we investigate and discuss numerical self-consistent
solutions of Eqs.~(\ref{BdGA}) and (\ref{BdGB}) with the HF potential
(the full version) and without it (the truncated version, by setting
$\Phi_{HF}(\rho)=0$ in the relevant expressions). Results are also
compared with a solution of Eqs.~(\ref{gapAnder}) and (\ref{HFAnder1}).
All the calculations are performed with the parameters typical for
aluminum~\cite{degen,fett}: $\hbar\omega_D=32.31~{\rm meV}$; $gN(0)=
0.18$, with $N(0)=m_ek_F/(2\pi^2\hbar^2)$ the bulk density of single-%
electron states at the Fermi level~[for $E_F$ see discussion after
Eq.~(\ref{ne})] and $k_F$ the bulk Fermi wavevector. For these
parameters the bulk BCS coherence length $\xi_0= 1.6~{\rm \mu m}$ is
significantly larger than the nanocylinder diameter. However, contrary
to the ordinary Ginzburg-Landau picture, the superconducting order
parameter now exhibits significant spatial variations in the transverse
direction due to the broken translational symmetry. The length of the
nanocylinder is taken as $L = 1~{\rm \mu m}\!\gg\!\lambda_F=2\pi/k_F$.
This is an optimal choice, upholding, on one side, the use of periodic
boundary conditions in the $z$ direction and, on the other side, it
results in a reasonable calculational time. As opposed to the truncated
BdG equations, their full version requires much more time for convergence
of the numerical procedure, and this time increases proportionally with
$L^2$. Numerically solving the Anderson equations (\ref{gapAnder}) and
(\ref{HFAnder1}) is less time-consuming and, so, we take $L = 5\,{\rm
\mu m}$ in this case.

In Fig.~\ref{fig1}(a) the spatially averaged order parameter
$$
\bar{\Delta} = \frac{2}{R^2}\int\limits_0^R {\rm d}\rho\,\rho \Delta(\rho),
$$
calculated from Eqs.~(\ref{BdGA}) and (\ref{BdGB}), is plotted in units
of the bulk order parameter ($\Delta_{\rm bulk}=0.25\,{\rm meV}$) versus
the nanocylinder diameter with and without the HF mean field. In Fig.~%
\ref{fig1}(b) the corresponding critical temperature $T_c$~(in units of
the bulk one) is given. As seen, both data-sets exhibit pronounced
size-dependent oscillations, typical of high-quality superconducting
nanofilms and nanowires with uniform thickness~\cite{film1,film2,sh1,sh2}.
Such oscillations result from single-electron subbands forming due
to the transverse quantization of the electron motion. With an
increase in the nanowire diameter, the subbands shift down in energy.
Each time when a new subband comes into the Debye window around the
Fermi level, the number of single-electron states contributing
to the superconducting order parameter increases, and a size-dependent
superconducting resonance develops. As follows from Fig.~\ref{fig1},
the quantum-size oscillations corresponding to the full version of the
BdG equations are somewhat shifted up. Hence, the difference between
the two sets of data is most significant for those diameters, where
a size-dependent superconducting resonance in the case without the
HF interaction is already present while in the full version such a
resonance only starts to develop. The difference is not so significant
but still survive when the resonance comes into its decay stage.
When the resonance is fully decayed (the off-resonant regime), the
HF corrections are practically negligible, and we arrive at the
situation similar to bulk. Notice that small differences between
the numerical results of the full and truncated BdG equations in the
off-resonant regime (due to beating patterns of the corresponding
curves), are because of the chosen nanowire length. Indeed, as follows
from calculations for several selected off-resonant diameters, such
beating patterns disappear when $L$ increases up to $20-30\,{\rm \mu
m}$, and the results with and without the HF interaction approach each
other.

Notice that maxima of $T_c/T_{c,{\rm bulk}}$ in Fig.~\ref{fig1}(b) are
generally higher than those of $\bar{\Delta}/\bar{\Delta}_{\rm bulk}$
in Fig.~\ref{fig1}(a). This is due to formation of new Andreev-type
states induced by the transverse quantum confinement (see details in
Ref.~\onlinecite{sh4}), which results in a decrease of $\bar{\Delta}/
(k_BT_c)$ below the bulk value $1.763$ at the resonant points. As seen
from Fig.~\ref{fig1}, inclusion of the HF interaction can slightly
reduce the resonant enhancements, with practically no effect on the
ratio $\bar{\Delta}/(k_B T_c)$.

%%%%%%%%%%%%%%%%%%%%%%%%%%%%%%%%%%%%%%%%%%%%%%%%%%%%%%%%%%%%%%%%%%%%%
\begin{figure}
\resizebox{0.8\columnwidth}{!}
{\rotatebox{0}{\includegraphics{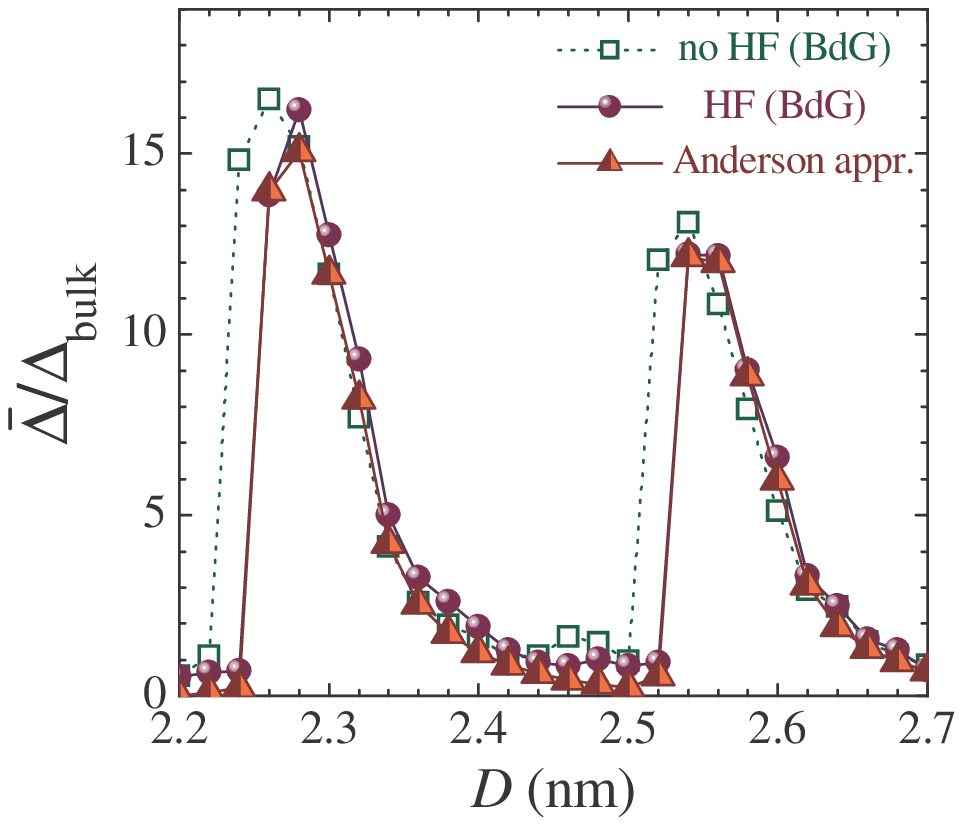}}}
\caption{$\bar{\Delta}/\Delta_{\rm bulk}$ versus the nanowire diameter
$(T=0)$: triangles correspond to the Anderson approximation [the HF field
is included]; stars and squares are the numerical results of the full and
truncated BdG equations, respectively.}
\label{fig3}
\end{figure}
%%%%%%%%%%%%%%%%%%%%%%%%%%%%%%%%%%%%%%%%%%%%%%%%%%%%%%%%%%%%%%%%%%%%%%%

%%%%%%%%%%%%%%%%%%%%%%%%%%%%%%%%%%%%%%%%%%%%%%%%%%%%%%%%%%%%%%%%%%%%%%%
\begin{figure}
\resizebox{0.8\columnwidth}{!}
{\rotatebox{0}{\includegraphics{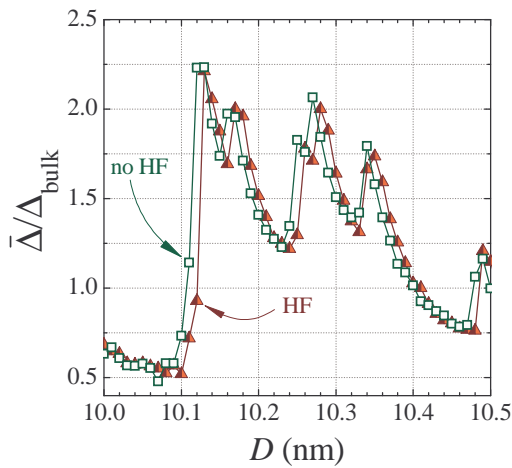}}}
\caption{HF versus no HF: $\bar{\Delta}/\Delta_{\rm bulk}$ as a function
of the nanowire diameter $D$ at zero temperature~[squares are the results
of the truncated BdG equations (no HF interaction); triangles are the
results of the Anderson approximation, see Eqs.~(\ref{gapAnder}) and
(\ref{HFAnder}).}
\label{fig4}
\end{figure}
%%%%%%%%%%%%%%%%%%%%%%%%%%%%%%%%%%%%%%%%%%%%%%%%%%%%%%%%%%%%%%%%%%%%%%%

In Fig.~\ref{fig2} we present different quantities calculated with the full
and truncated versions of Eqs.~(\ref{BdGA}) and (\ref{BdGB}) for two
diameters: the upper panel, for $D=2.24 \,{\rm nm}$; and the lower panel,
for $D=2.6\,{\rm nm}$. The upper panel represents the situation when a
superconducting resonance is developed for the truncated version but is not
yet present for the full version of the BdG equations. In Fig.~\ref{fig2}(a)
the superconducting order parameter $\Delta(\rho)$ calculated with (HF) and
without the HF interaction~(no HF, the inset), is plotted versus the
transverse coordinate $\rho/R$ for $T=0$. The spatial distribution of the
pair condensate is very different for these two cases: the data without
the HF interaction are larger by an order of magnitude, and even the
profile of $\Delta(\rho)$ is different. In Fig.~(\ref{fig2})(b) the
local electron density, i.e.,
\begin{equation}
n_e(\rho)=\frac{1}{\pi L}\sum_{jmk}\Bigl[u^2_{jmk}(\rho) f_{jmk} +
v^2_{jmk}(\rho)\bigl(1-f_{jmk}\bigr)\Bigr],
\label{neloc}
\end{equation}
is shown for the same diameter. As can be expected, now the difference
between the two data-sets is not so significant~(we keep to the same
value of the mean electron density $n_e$). Due to the attractive
character of the effective electron-electron interaction, the HF
potential forces electrons to go closer to the nanocylinder center.
However, the confining interaction has the major effect on $n_e(\rho)$
as compared to the HF potential producing only some small corrections.
From the results for the local electron density, it is possible to expect
that the single-electron wave functions are also not very sensitive to
the HF interaction. For nanowires, $|u_{\nu}({\bf r})|$ and $|v_{\nu}({\bf
r})|$ is nearly proportional to the corresponding single-electron wave
function~[see discussion above, after Eq.~(\ref{and})]. Hence, due to
Eq.~(\ref{norm}), the quantity $u^2_{jmk}(\rho) + v^2_{jmk}(\rho)$ can
provide us with the information about the single-electron distribution.
In Fig.~\ref{fig2}(c) $u^2_{jmk}(\rho) + v^2_{jmk}(\rho)$ is plotted
versus $\rho/R$ for the quantum numbers most sensitive to including the
HF interaction. We can indeed see that the effect of the HF potential
on the wave functions is minor. Similar conclusions can be obtained
from the lower panel of Fig.~\ref{fig2}. The only exception is that
the superconducting order parameter in Fig.~\ref{fig2}(d)~[$D=2.6\,{\rm
nm}$] does not change so much when including the HF potential. Notice
that $n_e(\rho)$ given in Fig.~\ref{fig2}(e) is practically the same
as in Fig.~\ref{fig2}(b). However, this is not true for $u^2_{jmk}(\rho)
+ v^2_{jmk}(\rho)$~[compare panel (c) with panel (f)]. The point is that
the integral $\int_{0}^{R}{\rm d}\rho\,\rho\,n_e(\rho) =n_e R^2/2$
changes with the radius but for the single-electron distribution
$u^2_{jmk}(\rho) + v^2_{jmk}(\rho)$ we have Eq.~(\ref{norm})].

From the results presented in Fig.~\ref{fig2}, one expects minor effects
on the single-electron wave functions due to the incorporation of the
HF interaction . This expectation can be put on a more solid ground by
using the Anderson approximation based on Eqs.~(\ref{gapAnder}) and
(\ref{HFAnder}). We remind that the Anderson approximation is quite good
for superconducting nanowires provided that it involves the true single-%
electron wave functions. Equations~(\ref{gapAnder}) and (\ref{HFAnder})
follow from Eq.~(\ref{and}) and, hence, as assumed, the single-electron
wave functions are not altered by our position-dependent HF interaction.
If this is a reasonable assumption, results of the Anderson approximation
constructed in this way, should be close to the results of the full BdG
equations. As seen from Fig.~\ref{fig3}, this is indeed the case. We can
conclude that the thickness-dependent shift of the superconducting
resonances in the presence of the HF interaction has nothing to do with
the single-electron wave functions. Its mechanism is due to the fact
that the position-dependent HF potential results in a change of the
single-electron energies measured from the Fermi level.

So far we considered extremely narrow nanowires, for the sake of simplicity.
However, a similar shift ($\approx 0.01 - 0.02\,{\rm nm}$) of the quantum-%
size oscillations due to the HF term survives until the total decay of the
quantum-size oscillations (up to diameters of about $50-70\,{\rm nm}$). In
particular, such a shift is clearly seen in Fig.~\ref{fig4}, where
numerical results of the truncated BdG equations for $D=10-10.5\,{\rm nm}$
are compared with a solution of Eqs.~(\ref{gapAnder}) and(\ref{HFAnder})
including the HF potential. Thus, we arrive at the following picture. In
the vicinity of a superconducting resonance, the bottom of some single-%
electron subband is situated close to the Fermi level. Therefore, a
repositioning of this subband with respect to the Fermi level can result
in a significant change of the number of single-electron states in the
Debye window and, so, in a remarkable increase/decrease of superconducting
characteristics. However, when bottoms of all single-electron subbands
are quite apart from the Fermi level, i.e., in the off-resonant regime,
a move of these subbands in energy produces much less important effect
on the number of single-electron states located in the Debye window.
This is why the decay of a superconducting resonance is accompanied by
a depletion of the influence of the HF potential.

\section{Conclusions}
\label{concl}

Quantum confinement breaks the translational symmetry in nanostructured
superconductors. In this case, despite the delta-function approximation
for the electron-electron interaction, the HF potential becomes position
dependent, and its contribution to the single-electron energy (measured
from the Fermi level) is a function of the relevant quantum numbers
(contrary to bulk!). By numerically solving the Bogoliubov-de Gennes
equations for a clean metallic nanocylinder, we have shown that such a
feature results in a shift of the curve representing the thickness-%
dependent oscillations of the critical temperature (the energy gap,
the order parameter etc.) to larger diameters. Though difference
between the numerical results with and without the HF potential can be
rather significant for some given diameters, the above mentioned shift
is less than typical metallic unit-cell dimensions and, so, can be
ignored. In particular, smoothing due to the thickness fluctuations
in real samples will shadow this effect. Thus, when solving the
Bogoliubov-de Gennes equations in the nano-regime, one is able to
neglect the HF term, which significantly simplifies a numerical
procedure.

\begin{acknowledgments}
Yajiang Chen thanks B. Xu and R. Geurt for helpful discussions. This
work was supported by the Flemish Science Foundation (FWO-Vl), the
Belgian Science Policy (IAP) and ESF-AQDJJ network.
\end{acknowledgments}


\begin{thebibliography}{99}
\bibitem{film1} Y. Guo, Y.-F. Zhang, X.-Y. Bao, T.-Z. Han, Z. Tang,
L.-X. Zhang, W.-G. Zhu, E.G. Wang, Q. Niu, Z. Q. Qiu, J.-F. Jia,
Z.-X. Zhao, and Q. K. Xue, Science {\bf 306}, 1915 (2004).

\bibitem{film2} M. M. \"Ozer, J. R. Thompson, and H.~H.~Weitering,
Nature Physics {\bf 2}, 173 (2006); {\it ibid.}, Phys. Rev. B {\bf
74}, 235427 (2006); M.~M.~\"{O}zer, Y.~Jia, Z.~Zhang, J.~R.~Thompson,
and H.~H.~Weitering, Science  {\bf 316}, 1594 (2007).

\bibitem{film3} D. Eom, S. Qin, M.-Y. Chou, and C. K. Shih, Phys.
Rev. Lett. {\bf 96}, 027005 (2006).

\bibitem{wire1} M. Savolainen, V. Touboltsev, P. Koppinen, K.-P.
Riikonen, and K. Arutyunov, Appl. Phys. A, {\bf 79}, 1769 (2004);
M. Zgirski, K.-P. Riikonen, V. Touboltsev, V., and K. Arutyunov,
Nano Lett. {\bf 5}, 1029 (2005); M. Zgirski, K. P. Riikonen,
V. Tuboltsev, P. P. Jalkanen, T. T. Hongisto, K. Y. Arutyunov,
Nanotechnology {\bf 19}, 055301 (2008).

\bibitem{wire2} M. L. Tian, J. G. Wang, J. S. Kurtz, Y. Liu,
M. H. W. Chan, T. S. Mayer, and T.~E. Mallouk, Phys. Rev. B
{\bf 71}, 104521 (2005).

\bibitem{wire3} L. Jankovi\v c, D. Gournis, P.~N.~Trikalitis,
I.~Arfaoui, T.~Cren, P.~Rudolf, M.-H.~Sage, T.~T.~M.~Palstra,
B. Kooi, J. De Hosson, M.~A. Karakassides, K. Dimos, A. Moukarika,
and T. Bakas, Nano Lett. {\bf 6}, 1131 (2006); N.~Tombros, L.~Buit,
I.~Arfaoui, T.~Tsoufis, D.~Gournis, P.~N.~Trikalitis, S.~J.~van der
Molen, P. Rudolf, and B. J. van Wees, Nano Lett. {\bf 8}, 3060 (2008).

\bibitem{wire4} F. Altomare, A. M. Chang, M. R. Melloch, Y. Hong,
and C. W. Tu, Phys. Rev. Lett. {\bf 97}, 017001 (2006).

\bibitem{sh1} A. A. Shanenko, M. D. Croitoru, and F. M. Peeters,
Europhys. Lett. {\bf 76}, 498 (2006); {\it ibid}, Phys. Rev. B
{\bf 75}, 014519 (2007).

\bibitem{sh2} A. A. Shanenko and M. D. Croitoru, Phys. Rev. B
{\bf 73}, 012510 (2006); A. A. Shanenko, M. D. Croitoru, M.
Zgirski, F. M. Peeters, and K. Arutyunov, Phys. Rev. B {\bf 74},
052502 (2006).

\bibitem{arut} K. Yu. Arutyunov, D. S. Golubev, and A. D. Zaikin,
Phys. Rep. {\bf 464}, 1 (2008).

\bibitem{han} J. E. Han and V. H. Crespi, Phys. Rev. B {\bf 69},
214526 (2004).

\bibitem{grig} I. Grigorenko, J.-X. Zhu, and A. Balatsky,
J. Phys.: Condens. Matter {\bf 20}, 195204 (2008).

\bibitem{fett} A. L. Fetter and J. D. Walecka, {\it Quantum
Theory of Many-Particle Systems} (Dover, New York, 2003).

\bibitem{degen} P. G. de Gennes, \emph{Superconductivity of
Metals and Alloys} (W. A. Benjamin, New York, 1966).

\bibitem{ander} P. W. Anderson, J. Phys. Chem. Solids
{\bf 11}, 26 (1959).

\bibitem{sh3} A. A. Shanenko, M. D. Croitoru, and F. M. Peeters,
Phys. Rev. B {\bf 78}, 024505 (2008); {\it ibid.}, Phys. Rev.
B {\bf 78}, 054505 (2008).

\bibitem{sh4} A. A. Shanenko, M. D. Croitoru, R. G. Mints, and
F. M. Peeters, Phys. Rev. Lett. {\bf 99}, 067007 (2007).

\end{thebibliography}
\end{document}